\documentclass[sigconf]{acmart}
\renewcommand\footnotetextcopyrightpermission[1]{}

\AtBeginDocument{%
  }

\setcopyright{acmlicensed}
\copyrightyear{2018}
\acmYear{2018}
\acmDOI{XXXXXXX.XXXXXXX}
\acmConference[Conference acronym 'XX]{Make sure to enter the correct
  conference title from your rights confirmation email}{June 03--05,
  2018}{Woodstock, NY}
\acmISBN{978-1-4503-XXXX-X/2018/06}

\begin{document}

\title{Bridging Psychological Safety and Skill Guidance: An Adaptive Robotic Interview Coach}

\author{Wanqi Zhang}
\email{Wzhang79@vols.utk.edu}
\orcid{0009-0007-2376-4942}

\affiliation{%
  \institution{University of Tennessee}
  \country{USA}
}

\author{Jiangen He}
\email{jiangen@utk.edu}
\affiliation{%
  \institution{University of Tennessee}
  \country{USA}
}

\author{Marielle Santos}
\email{msanto10@vols.utk.edu}
\affiliation{%
  \institution{University of Tennessee}
  \country{USA}
  }




\renewcommand{\shortauthors}{Trovato et al.}

\begin{abstract}
Social robots hold promise for reducing job interview anxiety, yet designing agents that provide both psychological safety and instructional guidance remains challenging. Through a three-phase iterative design study ($N=8$), we empirically mapped this tension. Phase~I revealed a ``Safety--Guidance Gap'': while a Person-Centered Therapy (PCT) robot established safety ($d=3.27$), users felt insufficiently coached. Phase~II identified a ``Scaffolding Paradox'': rigid feedback caused cognitive overload, while delayed feedback lacked specificity. In Phase~III, we resolved these tensions by developing an \textit{Agency-Driven Interaction Layer}. Synthesizing our empirical findings, we propose the \textbf{Adaptive Scaffolding Ecosystem}—a conceptual framework that redefines robotic coaching not as a static script, but as a dynamic balance between affective support and instructional challenge, mediated by user agency.
\end{abstract}



\begin{teaserfigure}
  \includegraphics[width=\textwidth]{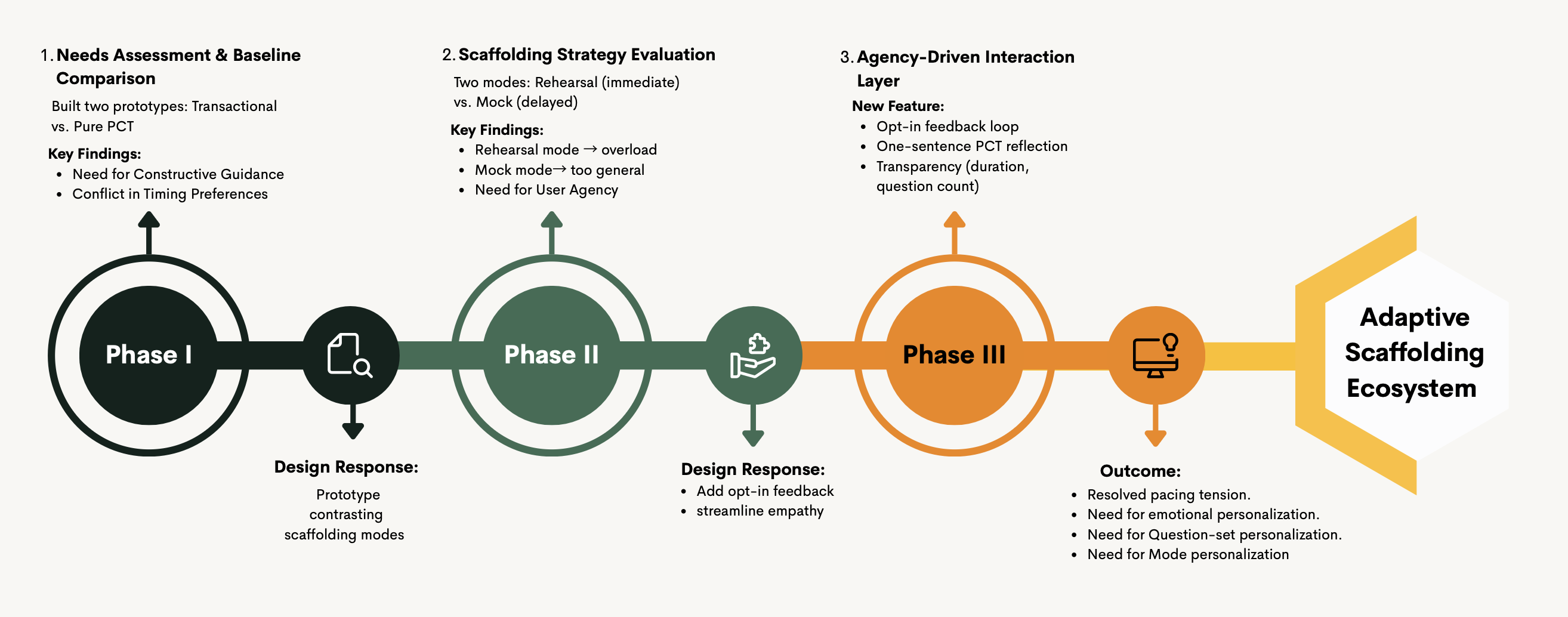}
  \caption{Three-phase developmental trajectory of the robotic coach, showing key findings and design changes culminating in the Adaptive Scaffolding Ecosystem.}
  \Description{flow-overview}
  \label{fig:teaser}
\end{teaserfigure}

\received{20 February 2007}
\received[revised]{12 March 2009}
\received[accepted]{5 June 2009}

\maketitle
\section{Introduction}
Reducing social anxiety and improving performance-related skills are 
central promises of social robots. While screen-based AI systems expand 
access to interview training~\cite{geng2024change,smith2014virtual}, 
physically embodied agents elicit a stronger sense of social 
presence~\cite{li2015benefit,david2014robot} and reduce perceived social 
pressure compared to humans~\cite{klkeczek2024robots,nomura2020people}. 
This combination positions social robots as ideal partners for 
high-stakes evaluative contexts such as job interviews, where anxiety 
can significantly undermine performance~\cite{powell2018meta}.

Existing interventions typically bifurcate into two approaches: 
skill-focused training (e.g., AI platforms analyzing non-verbal 
behavior~\cite{jadhav2024comprehensive}) or exposure-based therapy 
(e.g., VR simulations for desensitization~\cite{geng2024change}). 
However, few systems successfully integrate both. While 
cognitive-behavioral frameworks dominate robot-based 
interventions~\cite{rasouli2025co}, Person-Centered Therapy (PCT)--- 
emphasizing genuineness, unconditional positive regard, and em-
pathic understanding ---offers a powerful mechanism 
for establishing immediate psychological safety and mitigating acute anxiety~\cite{rogers1957necessary}. 
Yet, safety alone is insufficient for skill acquisition. Effective skill 
development requires instructional scaffolding: structured support that 
guides learners through their Zone of Proximal Development 
(ZPD)~\cite{vygotsky1978mind,wood1976role}.

A central challenge in Human--Robot Interaction (HRI) is redressing the 
imbalance between cognitive tutoring and affective support~\cite{woolf2009affect}. 
Prior educational research suggests that \emph{adaptive scaffolding}--- 
which dynamically adjusts support according to the learner's evolving 
understanding---outperforms fixed strategies in fostering 
self-regulated learning~\cite{azevedo2004does}. However, 
excessive external regulation can undermine learner 
agency~\cite{pekrun2006control}, creating a tension between 
providing guidance and preserving autonomy.

To explore this tension, we conducted a three-phase iterative design 
study ($N=8$). We assessed the ``Safety--Guidance Gap'' in a purely 
empathetic robot (Phase~I), identified a ``Scaffolding Paradox'' in 
conflicting feedback strategies (Phase~II), and resolved these tensions 
through an Agency-Driven Interaction Layer (Phase~III). This study 
addresses the following research questions:

\begin{itemize}
    \item \textbf{RQ1:} How can we design a PCT-based robot for establishing psychological safety in interview coaching?

    \item \textbf{RQ2:} How do different scaffolding strategies (immediate vs.\ delayed) affect cognitive load and perceived utility in robot-supported learning situations?
   
\end{itemize}

We make the following contributions:

\begin{itemize}
    \item \textbf{Preliminary evidence on the ``scaffolding paradox.''} 
    We find that while Rogerian empathy establishes safety 
    ($d = 3.27$), it lacks instructional utility. Conversely, rigid 
    feedback introduces cognitive costs, illustrating a practical 
    tension in HRI design.

    \item \textbf{A preliminary framework for adaptive scaffolding.} 
    We outline an ecosystem in which robotic coaches dynamically 
    calibrate affective tone and intervention intensity, mediated by 
    user agency, allowing systems to move between ``nurturer'' and 
    ``instructor'' roles as user needs evolve.
\end{itemize}

\section{System Implementation and Methodology}

\subsection{System Architecture}
To ensure ecological validity, we utilized the Misty II social robot to simulate physical presence. The robot was controlled via a custom web-based application (React.js/Node.js) integrating the OpenAI GPT-4 Realtime API. This architecture enabled low-latency, multimodal interactions, where system prompts guided verbal and non-verbal behaviors (e.g., nodding) over WebRTC. Researchers monitored system stability via a dashboard without intervening in the autonomous dialogue.

\subsection{Iterative Design Study Design}
We adopted a three-phase user-centred design process over Two weeks (see Figure~1). The study employed a **within-subjects design**, engaging the same cohort across all phases:
\textbf{Phase I (Baseline):} Comparing PCT principles against a transactional control.
\textbf{Phase II (Scaffolding Evaluation):} Testing conflicting feedback strategies (Rehearsal vs. Mock).
\textbf{Phase III (Validation):} Evaluating the final Agency-Driven system.

\textbf{Experimental Setup.} Sessions were conducted in a soundproof usability lab. Participants sat face-to-face with the robot (1m distance) to replicate professional interview dynamics. Researchers monitored conversation flow from an adjacent room via a one-way mirror to ensure system stability without disrupting the participant's psychological ``safe space.''

\subsection{Participants}
We recruited eight university students ($N=8$; 6 female, 2 male) aged 18--27 ($M = 21.25, SD = 3.24$) with self-reported moderate-to-high interview anxiety. Participants came from diverse backgrounds (e.g., Psychology, Nursing) and were active job seekers.
\subsection{Data Collection and Analysis}
We employed a mixed-methods approach targeting social perception, relational quality, and intervention efficacy.

\textbf{Quantitative Measures.} We utilized three instruments: (1) \textit{Robotic Social Attributes Scale (RoSAS-Mini)}: A concise measure capturing Warmth and Competence. (2) \textit{Barrett–Lennard Relationship Inventory (B–L RI Short Form)}: Operationalizes core PCT conditions—specifically empathic understanding, unconditional positive regard, and congruence—to quantify the “therapeutic alliance.” (3) \textit{Measure of Anxiety in Selection Interviews (MASI)}: Measured pre- and post-interaction levels using three subscales: Communication, Social, and Behavioral Anxiety.

\textbf{Qualitative Measures.} Semi-structured interviews followed each session to probe specific pain points (e.g., feedback timing, cognitive load). Transcripts were analyzed using reflexive thematic analysis.

\section{Iterative Design and Findings}

Our iterative process revealed a progression from establishing safety to navigating cognitive trade-offs, culminating in an agency-driven solution.

\subsection{Phase I: The Safety--Guidance Gap}

To address \textbf{RQ1}, we developed two contrasting system architectures using GPT-4 on the Misty II robot.
\begin{itemize}
    \item \textbf{Baseline (Transactional):} Simulating standard tools, the robot functioned as a neutral data collector, using static behaviors and brief acknowledgments (e.g., ``Understood'') to move through questions without emotional engagement.
    \item \textbf{PCT Condition (Empathetic):} Designed to replicate a counselor, the prompt enforced Rogerian principles (unconditional positive regard). Crucially, it included actuation instructions, transforming the LLM into a multimodal engine that matched emotional tone with gestures (e.g., warm nods, welcoming arm movements).
\end{itemize}

\subsubsection{Findings}
\textbf{Quantitative Validation.} Results confirmed that PCT successfully established psychological safety. On the RoSAS scale, the PCT condition achieved a Warmth score ($M=6.06, SD=0.73$) significantly higher than the Baseline ($M=2.06, SD=0.98$; $t(7)=-9.24, p<.001$). Similarly, participants reported a significantly higher therapeutic alliance on the B--L RI ($d=2.96, p<.001$). Competence ratings showed no significant difference ($p=.340$), suggesting warmth alone does not drive perceived capability.

\textbf{Qualitative: The Safety--Guidance Gap.} Despite the high safety, feedback revealed two critical design flaws:
\begin{itemize}
    \item \textbf{Need for Constructive Guidance:} Participants found the exclusive focus on positivity pedagogically limiting. 
    
    \item \textbf{Tension of Intervention Timing:} Participants were polarized regarding feedback timing. One group prioritized immediate analysis to capitalize on fresh memory, while others preferred deferring feedback to the end to preserve conversational flow.
\end{itemize}

This divergence highlighted that a static timing strategy could not meet conflicting cognitive needs, motivating the exploration of scaffolding strategies in Phase~II.

\subsection{Phase II: The Scaffolding Paradox}

Addressing the need for guidance (\textbf{RQ2}), Phase~II tested two conflicting scaffolding strategies with the same cohort ($N=8$, between-subjects):
\begin{itemize}
    \item \textbf{Rehearsal Mode (Strong Scaffolding):} Prioritized immediate instruction. The robot provided detailed professional feedback after \emph{every} response, creating a high-density ``perform--listen--learn'' structure.
    \item \textbf{Mock Mode (Weak Scaffolding):} Prioritized realism. The robot withheld feedback until a summative report at the end, preserving conversational flow.
\end{itemize}

\subsubsection{Findings: Critical Design Tensions}
The evaluation revealed a \textbf{Scaffolding Paradox}, where neither static extreme met user needs:

\begin{itemize}
    \item \textbf{Cognitive Overload (Rehearsal):} While the feedback was useful, the rigid pacing hindered processing. 
    \item \textbf{Dilemma of Authenticity:} The mandatory PCT reflection, when repeated mechanically alongside critique, damaged trust. 
    
    \item \textbf{The Specificity Gap and Need for Agency:} Conversely, Mock Mode felt too generic. Consequently, users rejected the binary choice, explicitly requesting the ability to regulate the flow .
\end{itemize}

These findings indicated that a static system---whether ``Strong'' or ``Weak''---was insufficient. Users required \textbf{Agency} (to control pacing), \textbf{Modulated Empathy} (to ensure genuineness), and \textbf{Operational Transparency} (to manage expectations).

\section{Phase III: The Interaction Layer: Agency-Driven Mode}
\subsection{The Final Version: The Agency-Driven Interaction Layer}
Addressing the Phase 2 trade-off between cognitive overload (strong scaffolding) and the specificity gap (weak scaffolding), we developed the \textbf{Agency-Driven Mode}. The system prompt was refined to operationalize control, authenticity, and transparency:

\begin{itemize}
    \item \textbf{Conditional Feedback Loop (Opt-in Valve):} To balance guidance with flow, the robot explicitly asked, ``Would you like professional feedback?'' after each response. ``Yes'' triggered a brief PCT reflection and concise critique; ``No'' skipped the evaluation to maintain momentum.
    \item \textbf{Streamlined Empathy:} We modulated PCT intensity by capping emotional reflections at one sentence. This kept the robot's empathy organic and professional rather than performative.
    \item \textbf{Operational Transparency:} The introduction established a clear mental model (e.g., ``8--10 questions, $\sim$20--30 minutes'') to mitigate the ``black box'' anxiety identified in Phase 2.
\end{itemize}

\subsection{Phase 3 Testing and Analysis}

\subsubsection{Findings: Agency as an Anxiety Buffer}
The Agency-Driven Mode achieved high usability validation. Qualitative analysis revealed that control over the interaction flow acted as a psychological buffer against anxiety. Notably, participants reported that the \textbf{ability to decline feedback} was as valuable as the feedback itself, successfully reframing the interaction from evaluation to collaboration.

\subsubsection{Emerging Design Tension: The Need for Deep Personalization}
While the Agency-Driven Mode resolved pacing issues, extended usage revealed a hierarchy of needs requiring deeper personalization across three levels:

\begin{itemize}
    \item \textbf{Emotional:} Users desired adaptable robot personalities (e.g., distinct interviewer profiles) rather than a fixed style.
    \item \textbf{Content:} Participants requested proficiency-adaptive content, distinguishing between general behavioral questions for novices and domain-specific inquiries for experts.
    \item \textbf{Mode:} Users envisioned structural flexibility beyond manual opt-ins, such as uninterrupted simulation modes for advanced practice versus supported modes for learning.
\end{itemize}

\vspace{0.5em}
\noindent \textbf{Conclusion for Part III.} The Agency-Driven Mode resolved the \textit{Scaffolding Paradox} by restoring user control. However, the demand for nuanced customization suggests that manual agency is merely a transitional step. These findings point toward an \textbf{Adaptive Scaffolding Ecosystem} that dynamically tailors personality and difficulty to the user's evolving state---a direction elaborated in the General Discussion.

\section{Evaluation of the Final System}

We evaluated the success of the final Agency-Driven Mode through comparative quantitative analysis and holistic user satisfaction ratings.

\subsection{Quantitative Validation: Efficacy \& Safety (Phase I vs. III)}

To validate the Agency-Driven ecosystem, we analyzed its impact on anxiety reduction (Efficacy) and compared social attributes against the purely empathetic baseline (Safety).

\textbf{1. Intervention Efficacy (MASI Results).}
A series of paired-samples $t$-tests on pre- vs. post-interaction MASI scores revealed that the system successfully reduced acute anxiety.
\begin{itemize}
    \item \textbf{Social Anxiety:} Showed the strongest reduction, dropping significantly from $M=3.31$ ($SD=0.39$) to $M=2.65$ ($SD=0.45$; $t(7)=7.08, p<.001$).
    \item \textbf{Communication Anxiety:} Significantly decreased from $M=2.99$ ($SD=0.73$) to $M=2.56$ ($SD=0.61$; $t(7)=2.41, p=.047$).
    \item \textbf{Behavioral Anxiety:} Did not change significantly ($p=.685$). This suggests that while the robot effectively alleviates cognitive and social distress, altering deep-seated physiological responses may require longer-term exposure.
\end{itemize}

\textbf{2. Preservation of Safety }
A critical design concern was whether introducing instructional critique (Phase~III) would erode the psychological safety established by the pure counselor (Phase~I). Comparative analysis confirmed that the therapeutic bond was sustained:
\begin{itemize}
    \item \textbf{Sustained Warmth:} There was no significant difference in RoSAS Warmth ($p = .425$) or B--L RI Relationship Quality ($p = .073$) between the verbose Pure PCT and the streamlined Agency-Driven Mode.
    \item \textbf{Enhanced Competence:} The Agency-Driven Mode achieved the highest numerical Competence scores ($M=6.25$), surpassing the Phase~I baseline ($M=5.81$).
    \item \textbf{Low Discomfort:} Discomfort ratings in Phase~III ($M = 1.75$) remained significantly lower than the Transactional Control ($p<.05$) and comparable to Phase~I ($p=.417$), indicating that the ``Opt-in'' mechanism effectively mitigated evaluation anxiety.
\end{itemize}

\textit{Synthesis:} These results demonstrate a successful dual outcome: the system effectively reduced user anxiety (MASI) while maintaining the high levels of warmth and trust (RoSAS/B-L RI) typically reserved for PCT agents.

\subsection{System Effectiveness and Satisfaction}
Participants rated the final system's holistic impact on ten dimensions (0--100 scale). The results highlight three key strengths:

\begin{itemize}
    \item \textbf{High Utility:} Participants rated ``Information usefulness'' ($M = 91.3, SD = 11.3$) and ``Support for stress management'' ($M = 92.5, SD = 11.7$) as the highest attributes. Consequently, retention intent was high (``Will continue practicing,'' $M = 92.5$).
    \item \textbf{Confidence Building:} Despite the high-stakes context, users reported substantial ``Increased confidence'' ($M = 83.8, SD = 19.2$), verifying that agency-driven feedback supported self-efficacy.
    \item \textbf{Acceptable Artificiality:} Lower scores for ``Natural behavior'' ($M = 65.0$) reflect the inherent artificiality of robotic role-play. Importantly, participants nonetheless felt ``Ready for a human interview'' ($M = 76.3$), validating the system's role as a preparatory tool rather than a human replacement.
\end{itemize}

\section{Discussion}

Our findings offer empirical evidence for bridging the gap between emotional support and instructional guidance in HRI. We discuss two key implications: the role of agency as a scaffolding strategy and the theoretical shift toward adaptive ecosystems.

\subsection{Agency as a Scaffolding Strategy}
Current educational robots often treat ``user control'' merely as a UI feature. However, our Phase~III results suggest that in high-stakes, anxiety-inducing contexts, \textbf{agency itself is a scaffolding strategy}.
By allowing users to toggle between ``Learning Mode'' (Feedback On) and ``Flow Mode'' (Feedback Off), the Opt-in mechanism enabled users to self-regulate their exposure to critique. This aligns with Self-Determination Theory: preserving autonomy buffers the social threat usually associated with evaluation~\cite{ryan2000self}. Unlike Phase~II's rigid pacing, which forced users into a Zone of Proximal Development (ZPD) they were not emotionally ready for, the Agency-Driven layer allowed users to enter the ZPD on their own terms.

\subsection{Toward an Adaptive Scaffolding Ecosystem}
While the \textit{Manual Agency} (Opt-in) proved effective, qualitative feedback (e.g., decision fatigue) points to the necessity of \textit{Intelligent Adaptation}. Participants eventually requested a system ``smart enough to know when to step in'' (P8).

We propose a theoretical shift from manual choices to an \textbf{Adaptive Scaffolding Ecosystem}.
\begin{itemize}
    \item \textbf{Theoretical Basis:} As \citet{azevedo2004does} demonstrated, adaptive scaffolding—which dynamically adjusts to the learner's emerging understanding—is significantly more effective at facilitating self-regulated learning than fixed scaffolding (like our Phase~II). Furthermore, \citet{woolf2009affect} emphasize that intelligent tutors must ``redress the cognitive versus affective imbalance.''
    \item \textbf{Future Architecture:} Drawing on this, future robotic coaches should not rely on a static persona. Instead, they should utilize \textbf{implicit sensing} (e.g., analyzing speech latency, prosody, or hesitation) to infer anxiety states in real-time. Based on this data, the system should autonomously slide along two axes—\textit{Affective Tone} and \textit{Intervention Intensity}—shifting between a ``Nurturer'' role to reduce anxiety and an ``Instructor'' role to build skills, without requiring explicit user commands.
\end{itemize}

\subsection{Limitations}
Our study is limited by a small sample size ($N=8$) and a short duration, which may introduce novelty effects. Additionally, our evaluation relied on self-reported metrics (RoSAS, B--L RI); future work should incorporate objective performance benchmarks (e.g., mock interview scores graded by HR experts) and physiological data (e.g., heart rate variability) to verify skill acquisition and stress reduction objectively.

\section{Conclusion}

This study addressed the fundamental tension between psychological safety and instructional guidance in robotic coaching. Through a three-phase iterative design, we identified a \textbf{Scaffolding Paradox}: rigid guidance overloads the user, while pure empathy leaves them unskilled.

We contribute an \textbf{Agency-Driven Interaction Layer} that resolves this paradox. By integrating a conditional feedback loop with streamlined empathy, our system achieved high user satisfaction ($91.3/100$ utility) without sacrificing the therapeutic bond established by PCT ($p>.05$ for Warmth retention). Ultimately, this work suggests that the next generation of AI coaches must evolve to \textbf{adaptive ecosystems} that attune to the learner's shifting cognitive and emotional needs, bridging the divide between the comfort of a counselor and the rigor of a coach.


\bibliographystyle{ACM-Reference-Format}
\bibliography{sample-base}

\appendix

\end{document}